\documentclass[11pt,twoside]{atmp}

\usepackage{amsmath,amssymb}

\usepackage[all]{xy}
\usepackage{color}
\numberwithin{equation}{section}

\copyrightnotice{2007}{1}{1}{23} %% year, volume, first page, last page.
\begin{document}

\title[A New Formulation of General
Relativity] {A New Formulation of General Relativity - Part I:
Pre-Radar Charts as Generating Functions for Metric and Velocity}

\arxurl{<hepreference}

\author[ Joachim Schr\"oter]{ Joachim Schr\"oter}

\address{Department of Physics, University of Paderborn,\\ D-33095 Paderborn, Germany}  %lines should be separated with double backslashes: \\
\addressemail{J@schroe.de}

\begin{abstract}
{\small{ In this paper (Part I of a series of three papers) an
axiomatic formulation of GR is given, here use is made of the
concept of pre-radar charts.  These charts have "infinitesimally"
the same properties as the true radar charts used in space-time
theory. Their existence in GR has far-reaching consequences which
are discussed throughout the papers. For the sake of simplicity
and convenience I consider only such material systems the state of
which is defined by a velocity field, a mass density and a
temperature field. But the main results hold also for more complex
systems. It follows from the axiomatics that the pre-radar charts
define an atlas for the space-time manifold and that, in addition,
they generate the metric, the velocity field and the displacement
of the matter. Therefore, they are called generating functions.
They act like "potentials". At the end of the paper it is shown
that the existence of pre-radar charts allows to simplify the
original axiomatics drasticly. But the two versions of GR are
equivalent.}}
\end{abstract}

\maketitle

\cutpage

\section{Introduction}

{\bf 1.1:} The subject of this treatise is the general theory of
relativity (GR) in its classical form. In a strict sense, GR is
not a single physical theory, rather it is a class of theories
which share common features. From this point of view Schwarzschild
space-time, Robertson-Walker space-times etc. are counted as separate relativistic theories. \\
The common features of all these theories are ''principles'' they
obey. More properly these principles should be called ''axioms''
because they have the same status as the axioms in mathematical
theories. An inspection of relativistic theories reveals that
their axioms can be grouped into five classes as follows.

\begin{itemize}
\item[\bf GK:] {Axioms concerning geometry and kinematics.}
\item[\bf EM:]{Axioms concerning matter and its motion, e.g. equations of motion and constitutive equations.}
\item[\bf ED:]{Axioms concerning electromagnetism, e.g. Maxwell´s equations and constitutive equations.}
\item[\bf EE:]{Einstein equation and constitutive equations.}
\item[\bf AC:] {Additional conditions, e.g. initial conditions.}
\end{itemize}

Clearly, in vacuum theories the axioms EM and ED are empty. The
above classification of axioms possibly includes redundancies.\\

{\bf 1.2:} Among the many forms of presenting physical theories
there are two extreme forms which are of special interest in
our context. \\
(i) The first form is characterized by the property that the
fundamental terms as in our case a set of events $M$, an atlas $
\mathcal{A}$, a metric $g$, a velocity field etc. are implicitly
determined solely by axioms. Examples of such formulations of
theories are well known
in many branches of physics. \\
(ii) The second form of a physical theory can be characterized as
''model theory'', also widely known under the label ''solution''.
In this case the letters  $M$, $\mathcal{A}$, $g$ etc. are
replaced by explicit terms of mathematical analysis, and the
axioms of case (i) occur as theorems, i.e. the axioms are
satisfied by
these explicit terms. Also this type of a theory is well known in physics.\\
Clearly, mixed forms are on the market, too. In what follows, I
will consider relativistic theories according to the first form.
Formulating their axioms, I will make use of some results of the
space-time theory (STT) which is developed in \cite{schroe87},
\cite{schroe92}, \cite{schelb92} and which is reviewed in
\cite{schroe93}. A detailed account of this STT can be found in
\cite{schelb97}. More specific, I will take some features of radar
coordinates in order to define a weaker form of them which I call
pre-radar coordinates. Using these coordinates in the context of
GR is the essential new aspect of this treatise. They can be
comprised into one function $\Psi$ depending on two events $p,q$
which is a generating function for the atlas $\mathcal A$, the
metric $g$ and the velocity $v$, and which, in addition,
determines the
integral curves of $v$.\\

{\bf 1.3:} A few remarks may illustrate the notion of radar
coordinates as it is used in \cite{schroe87} to \cite{schelb97}.
Let $A$ be an observer, $b$ an event and $ t_{1}, t_{2}$ times
measured by the clock of $A$. Finally, let $e = (e^1, e^2, e^3)$
be the direction of a light signal which leaves $A$ at instant $
t_{1}$ and comes back to $A$ at $ t_{2}$ after being reflected at
$b$. Then, if

\begin{equation*}
t := \frac{1}{2} (t_{2} + t_{1}), \quad r :=
\frac{1}{2}(t_{2}-t_{1})
\end{equation*}

the radar coordinates $ x$ of $b$ are given by $x= (re^{1},
re^{2}, re^{3}, t)$. Here
and in the sequel all quantities are dimensionless, and $c = 1$.\\

If $A$ describes this situation with the help of its own radar
coordinates he or she will get in two dimensions of
$\mathbb{R}^{4} $ the picture shown in Fig.~\ref{fig1}. Here the
outgoing light signal is a straight line by definition, whereas
the incoming signal is not necessarily straight, but it is a curve
which is a subset of a Minkowskian backward light cone.
Intuitively, a pre-radar chart of an observer $A$ has the
properties of a radar chart only in an "infinitesimal
neighborhood" of the worldline of $A$. In each case, radar
coordinates are also pre-radar
coordinates as introduced in Definition (3.2) of Section 3.1.\\
It is to be emphasized that the term radar coordinate is not
uniformly used in literature. Two examples may illustrate it.
Coleman and Kort\'{e} \cite{Coleman80} define radar charts which
are similar to those introduced above. The difference is that the
measurement of the direction $e$ is performed on the incoming part
of the radar signal. The radar coodinates used in EPS axiomatics
\cite{Ehlers72} are more different from ours. Here the radar
coordinates $(u, v, u', v')$ of an event $a$ are defined by two
observers $A$ and $A'$ which send out radar signals starting at
the times $u, u'$ and arriving at $v, v'$ after being reflected at
$a$.

\begin{figure}[htb]
\centerline{\unitlength.9cm%
\begin{picture}(7.5,10)
\thicklines%
\put(2,0.5){\vector(0,1){9}} \put(2,1.5){\line(1,1){3.5}}
\put(2,5){\line(1,0){3.5}}
\multiput(2,8.5)(.62,-.62){6}{\line(1,-1){.4}}
\put(1.6,1.4){$t_1$} \put(1.6,8.4){$t_2$} \put(1.7,4.9){$t$}
\put(2.3,9.2){$A$} \put(5.7,4.9){$b$}
\thinlines%
\put(2.1,1.95){$e$} \qbezier(2,2.3)(2.4,2.3)(2.5,2.0)
\end{picture}}
\begin{center}Figure 1\label{fig1}\end{center}
\end{figure}

{\bf 1.4:} Though the class of theories I want to consider are
continuum theories the notions observer, particle or real point
are employed. This does not contradict the continuum point of
view. Rather it reflects only the fact that we have two
possibilities to describe continuum systems, namely in the way of
Lagrange as systems of particles or in the way of Euler by fields.
Later on I will use a mixture of both these descriptions.

{\bf 1.5:} Throughout this paper I will use the following
strategy: each material point, i.e. each point which contributes
to gravitation, is a part of the system of pre-radar observers.
However it is possible that there are gravitationally irrelevant
test particles which are pre-radar observers.

{\bf 1.6:} As indicated in Section 1.2 I intend to give a new
axiomatic formulation of General Relativity. There is already a
considerable amount of work done in axiomatics of relativity and
space-time theory which cannot be reviewed here. Rather I refer to
the paper of Schelb \cite{schelb97} where the reader will find an
almost complete list of relevant papers. One of the most recent
treatises in this field is that of Hehl and Obukhov \cite{Hehl05}
where a Lorentz metric on a manifold is constructed using
electrodynamics without metric. But in what follows I do not adopt
this general supposition, rather I will
 take into account only some special features of electrodynamics
by using light signals as a basic concept. (Cf. also
\cite{Andre04} and the literature quoted there.)

\section{Description of the Systems to be Considered}

{\bf 2.1:} In what follows, I will not treat the most general
continuum systems, rather I consider only material systems which
can be described by one velocity field $v$, one mass density
$\eta$ and one empirical temperature $\vartheta$. This means that
mixtures of fluids, especially plasmas are not taken into account.
(Mixtures of fluids are treated e.g.\ in \cite{neugeb80}, p.\
131).The above restriction is only a matter of convenience because
the main results of this treatise are also valid
for more complex systems. \\

Though each system considered is supposed not to have
electromagnetic interaction it is assumed that all observers,
i.e.\ particles of the material system and test particles can
exchange light signals. It is assumed that these signals are
irrelevant with respect to any kind of interaction. The only thing
they can transport is information. They have the same status as
test particles. The reason to take into account light signals is
that we want to introduce pre-radar-coordinates. This can be
effected
most simply if we have light signals at our disposal. \\

{\bf 2.2:} In order to formulate a physical theory in the sense of
(i) of Section 1.2 which is adequate to treat the systems
described in Section 2.1, the fundamental mathematical terms have
to be specified with the help of which the whole theory can be
formulated and which are implicitly determined by the axioms of
the theory. There are two kinds of fundamental terms, the
so-called \textit{base sets} and the so-called \textit{structural
terms}.  The first ones contain the signs for the objects to be
treated, whereas the latter, the structural terms define the basic
properties of these objects. Expressed mathematically, the
structural terms are relations which are elements of sets
constructed from the base sets solely with the help of the
operations
"power set" and "Cartesian product".\\

Let us first specify the base sets. The most fundamental term in
any relativistic theory is the set of signs for events. This is
indispensable! In addition, we want to speak about particles some
of which bear gravitationally active masses, and we take into
account (light) signals. Finally, we need the real numbers because
we have to express the values of some quantities by numbers. \\

Let us now come to the \textit{structural terms}. They are the
metric, the velocity field, the mass density and the (empirical)
temperature field. As usual in GR  the set of events should be a
manifold. This means that there are coordinates defined on certain
sets of events. The structural term which is introduced in the
present context is a relation  which assigns four coordinates to
each event in
some neighborhood of an observer.\\

It is convenient to assign a common name to the theories treated in this paper.\\
\textit{\bf Notation (2.1):} The physical theories which are
determined by the above mentioned (and in Section 2.3 precisely
described) base sets and structural terms which in turn are ruled
by the axioms of Sections 3 and 4 are denoted $ \Phi_{R} $. Since
$ \Phi_{R} $ represents a class of physical theories (in a
strict sense) it is called a frame theory. \\

{\bf 2.3:} Summing up the considerations of Section 2.2 we arrive
at the following result:
The {\it base sets} of $ \Phi_{R} $ are $M, P, S$, $ \mathbb{R}$:\\
$M$ is the set of signs $a$, $b$, ... etc. for events \\
$P$ is the set of signs (i.e. indices) $A, B$, ... etc. for particles \\
$S$ is the set of signs (i.e. indices) $s, s'$, ... etc for signals \\
$ \mathbb{R}$ is the set of the real numbers as usual.

The {\it{structural terms}} of $ \Phi_{R} $ are  $ \hat{ \Psi}, g, v, \eta, \vartheta:$ \\
$\hat{ \Psi}$ determines the pre-radar coordinates, i.e. $ (A, b,
x) \in \hat{\Psi}$ means that observer $A$ coordinatizes event $b$
by $x$

\noindent
$ g $ is the metric \\
$ v$ is the velocity field \\
$ \eta$ is the mass density \\
$ \vartheta $ is the temperature field.

\section{Geometric and Kinematic Axioms}

Following the notation of Section 1.1 the axioms of this section
are denoted GK. Throughout the Sections 3, 4, and 5 the same
natural numbers $ \textit{k} \;\text{resp.}\; r= \textit{k}-1$ are
meant in phrases like  "$ \ldots \in C^k, k \ge 3 $" or "$\ldots
\in C^r, r \ge 2$".

\subsection{Pre-radar charts}

Intuitively, a pre-radar chart is a chart which has some (or
perhaps all) of the properties of a true radar chart. Therefore,
the following axioms can be motivated by pointing to the fact that
true radar charts have a certain property. In Section 1.3 it was
outlined how a radar observer coordinatizes a neighborhood of its
worldline: he or she needs clocks and devices for measuring
directions. In what follows it is assumed that each observer uses
one and only one clock and one and only one directional measuring
instrument. Therefore, the radar charts of observers thus equipped
are unique. Hence the following axiom is self-evident: \\

\textit{\bf GK 1.1}: The structural term $ \hat{\Psi} $ is a
function: $\hat{\Psi} : \bigcup_{A \in P} \{ A \} \times V_{A}
\rightarrow
\mathbb{R}^{4},$ where $ V_{A} \subset M $ and $ V_{A} \neq \varnothing $. \\

It is useful to introduce some notation. \\
{\bf Definition (3.1)}: For short we write $ \psi_{A} :=
\hat{\Psi} (A,\cdot) $ and $ {\mathcal O}_{A} := \mathrm{ran}
\psi_{A}$; by definition $\mathrm{dom}  \psi_{A}=V_{A} $.
Then let ${\mathcal A} = \{ ( V_{A}, \psi_{A}) : A \in P \}.$\\

The next axiom expresses that $ \hat{\Psi}$ determines a manifold structure on $ M $. \\

\textit{\bf GK 1.2:} $ {\mathcal A} $ is a $ C^{\textit{k}}
$-atlas, $ \textit{k} \geq 3 $ such that $(M, {\mathcal A}) $ is a
connected Hausdorff
manifold.  \\

A motivation for further axioms comes from the fact that each
radar observer $A$ coordinatizes himself or herself by $
(0,0,0,t)$, the parameter $t$ being the time $A$ measures. In
addition, the fourth
component of each quadruple of radar coordinates measured by $A$ is a time at $A$. \\

\textit{\bf GK 1.3:} For each $ A \in P $ there are two real
numbers $ u_{1}, u_{2} $ with $ -\infty \leq u_{1} < u_{2} \leq
\infty $ such that for each $ \tau \in ] u_{1}, u_{2} [ = :J_{A}$
the relation $ (0,0,0,\tau) \in \mathcal{O}_{A}$ holds; moreover,
if $ y_{1} < u_{1}$ or $ u_{2} < y_{2}$, then $ \{ (0,0,0,
\rho ): \rho \in ] y_{1}, y_{2} [ \;\} \not\subset {\mathcal O }_{A}$, i.e. $J_{A}$ is maximal.  \\

Finally, some additional notation is introduced by

\textit{\bf Definition (3.2)}: 1.\ The term $ \mathcal{D}$ is the
differential structure containing all charts which are $
C^{\textit{k}}$-compatible, $ \textit{k} \geq 3 $ with $ {\mathcal
A} $.\\
2. The coordinates determined by the charts $(V_A, \psi_A)$ of $
\mathcal{A}$ will be called for short $ A$-coordinates etc.\,
whereas the others are denoted by their coordinate functions $ \chi $ etc.\\
3. Within the theory $ \Phi_{R} $ the charts of $ {\mathcal A}$ are called pre-radar charts. \\

 \subsection{Worldlines of particles}

 In our context there are two possibilities to define the worldline of an  observer. First, the worldline of $A$ is the set of events which
 occur at $A$. Second, the worldline of $A$ is the set of events  which $A$ coordinatizes by $ (0,0,0,t)$. This leads to the  following \\
\textit{\bf Definition (3.3)}: For each $ A \in P $ the
(surjective) function $\gamma_{A} : J_{A} \rightarrow W_{A}
\subset M$ is defined by $ \gamma_{A} (t)= \psi_{A}^{-1}
(0,0,0,t), t \in J_{A} $. The set $ W_{A} : = \mathrm{ran}
\gamma_{A} $ is called the
worldline  of $A$. As usual $ \dot{\gamma}_{A}$ denotes the velocity of $A$. \\

\textit{\bf Definition (3.4)}: For each $A \in P $ the clock $
{\mathcal U}_{A} $ of $A$ is defined by ${\mathcal U}_{A} (a) = \psi_{A}^{4} (a), a \in W_{A} $. \\

\textit{\bf Remark (3.5)}: 1. From Definition (3.3) it follows
that $\psi_{A} \circ \gamma_{A} (t) = (0,0,0,t)$.
Therefore, the function $ \gamma_{A} $ is of class $ C^{\textit{k}}, \textit{k} \geq 3 $ and bijective. \\
2. From Definition (3.4) one concludes that

\begin{equation*}
{\mathcal U}_{A}(\gamma_{A} (t)) = \psi_{A}^{4} \circ
\psi_{A}^{-1} (0,0,0,t) = t = \gamma_{A}^{-1} (\gamma_{A} (t)).
\end{equation*}

Hence ${\mathcal U}_{A} (a) = \gamma^{-1}_A (a) \; \text{for all} \; a \in W_{A}$.\\

Since we want to describe a continuum, the set of particles should
be a continuum. In addition, we want to describe the system  by a
smooth velocity field. Therefore, the worldlines of different
particles cannot have common events. Hence the next two
axioms are self-evident. \\

\textit{\bf GK 2.1}: The cardinality of $P$ is that of a continuum
and $ \bigcup_{A \in P} W_{A} = M $. \\

\textit{\bf GK 2.2}: For all $ A, B  \in P$: if $ A \neq B$ then $
W_{A} \cap W_{B} = \varnothing $.

In other words this axiom expresses that the set $P$ of particles
represents a congruence.

\textit{\bf Remark (3.6):} It follows directly from the Axioms GK
2.1 and 2.2 that there is a function $ F : M \rightarrow P $ which
is surjective and which is given by $ F(a) = A$ for all $ a \in
W_{A}: =\; \mathrm{ran}\; \gamma_{A} $.

With the help of $ F $ we are now able to define a function $ \Psi
$ which later on turns out to be a generating function for the
metric $ g$ and the velocity $ v $. \\

\textit{\bf Definition (3.7)}: The function $\Psi : \bigcup_{q \in
M} V_{F(q)} \times \{ q \} \rightarrow \mathbb{R}^{4}$ is defined
by $ \Psi (p,q) = \hat{\Psi} (F (q), p)  = \psi_A (p)$ for  $A =
F(q).$ The values of $\Psi$ and $\hat{\Psi}$ are mostly written as
row vectors: $\Psi = (\Psi^1, \cdots, \Psi^4)$ etc.\ . But
occasionally it is more convenient to
write them as column vectors: $\Psi = (\Psi^1, \cdots, \Psi^4)^T$ etc.\ . \\

At this point, $\Psi$ is nothing but another form of $ \hat{\Psi}
$ which has the advantage that it allows to express the
intuitively reasonable property of pre-radar coordinates that
''neighboring'' observers attribute ''neighboring'' coordinates to
the same event. This means that $ \Psi $ is ''smooth'' with
respect to all of its arguments. Therefore, the
following axiom is natural. \\

\textit{\bf GK 2.3}: $ \Psi $ is of class $ C^{\textit{k}} $ with $ \textit{k} \geq 3 $. \\

\subsection{Axioms for the metric}

Since the metric $g$ is defined in the usual way it suffices to
write down the axioms governing $g$. \\

\textit{\bf GK 3.1}: The structural term $g$ is a function $g :
\bigcup_{a \in M } \{a \} \times (T_{a} M \times T_{a}M)
\rightarrow
\mathbb{R}$, where $ T_{a} M $ is the tangent space at $ a \in M $. \\

It is convenient to introduce the\\
\textit{\bf Notation (3.8)}: $ g (a, w, w^{ \prime }) = g (a) (w, w^{ \prime}) $ and $ g (a) = g_{a} = g (a) (\cdot, \cdot )$. \\

\textit{\bf GK 3.2}: $ g = g ( \cdot) $ is a ${0\choose{2}}$-tensor field of class $C^r, r  \geq 2$.\\

\textit{\bf GK 3.3}: For each $ a \in M : g_a $ is symmetric, non-degenerate and of signature 2. \\

With the help of $g$ the clocks used by the observers now can be
specified: they are to show proper time. This is the content of
the next axiom which joins particles and metric. \\

\textit{\bf GK 3.4:} For each $ A \in P $ and each $ t \in J_{A}
$: $g ( \gamma_{A} (t)) ( \dot{\gamma}_{A} (t), \dot{\gamma}_{A}
(t)) = -1$.

\subsection{Properties of signals}

In this subsection the relation between particles and signals is
studied. The motivation for the following axioms comes from
space-time theory (cf.
\cite{schroe87},\cite{schroe92},\cite{schelb92},\cite{schroe93},\cite{schelb97})
where proper radar charts are treated in great detail. It is not
possible to repeat all arguments of these papers here. One result
may suffice. A true radar observer $A$ describes outgoing light
signals within his or her coordinate system by straight lines
leaving the worldline of $A$ by an angle of 45 degrees. In what
follows one should also
have in mind that $S$ is merely a set of indices for signals. \\

\textit{\bf GK 4.1}: For each $ s \in S $ there is a function
$\sigma_{s} : K_{s} \rightarrow  W^\ast_s \subset M, \; K_{s}
\subset  \mathbb{R}$ with the following properties:

$ \sigma_{s}$ is of class $ C^{\textit{k}}, \textit{k} \geq 3 $ and a null geodesic; moreover there is an $ A \in P $ such that \\

\setlength{\parindent}{1cm}{
$K_{s} = [ t_{s0}, t_{s1}] \subset J_{A};$\\

$\psi_{A}^{4} ( \sigma_{s} (t) ) = t  \quad \text{for all} \quad  t \in K_{s}; $\\

there is an $ a \in W_{A}$ such that $ \sigma_{s} (t_{s0}) = a$;\\
\indent there is an $e = ( e^{1}, e^{2}, e^{3}) \in S^{2} $ (the 2-sphere) such that\\

$\frac{d}{dt} \psi_{A} ( \sigma_{s} (t_{s0})) = (e^{1}, e^{2}, e^{3}, 1)$.}\\

\setlength{\parindent}{0cm}
To a certain extent also the converse of this axiom is needed. \\

\textit{\bf GK 4.2}: For each $A \in P $, for each $ a \in W_{A} $
and for each $(e^{1}, e^{2}, e^{3}) \in S^{2}$ there is an $ s \in
S $ such that the function $ \sigma_{s} $
determined by axiom GK 4.1 has the following properties: \\

\setlength{\parindent}{1cm}
$\sigma_{s} (t_{s0}) = a ;$\\

$\frac{d}{dt} \psi_{A} ( \sigma_{s} (t_{s0})) = ( e^{1}, e^{2},
e^{3}, 1)$. \setlength{\parindent}{0cm}

\subsection{Velocity}

According to Definition (3.3) the velocity of particle $A$ is
$\dot{\gamma}_{A}$. Therefore, the field $ v $ is determined by $
\dot{\gamma}_{A}$. This is the content of Axiom GK 5. \\

\textit{\bf GK 5}: The structural term $ v $ is a function $v : M
\rightarrow T M$ which is defined for each $ b \in M $ by $v (b) =
\dot{\gamma}_{F(b)} ( \gamma_{F(b)}^{-1} (b))$ and which is of
class $ C^{r}, r \geq 2$.

\section{Further Axioms}

\subsection{Equations of motion}

Up to now only the geometric and the kinematic part of the theory
$\Phi_{R} $ was treated, and this part can be characterized,
roughly speaking, by the phrase: everything is smooth. For the
formulation of the further axioms we have to take matter into
account. The basic quantities describing matter can again be
smooth, but there are many interesting systems which show material
discontinuities. It is not possible to treat these different classes of systems by the same kind of axioms. \\
Therefore I restrict the further studies to the most simple class
of systems which are those with smooth $ \eta $ and $ \vartheta $. In this case the following axiom is obvious. \\

\textit{\bf EM1}: The structural terms $ \eta $ and $ \vartheta $
are functions $\eta : M \rightarrow \mathbb{R}, \quad   \vartheta
: M \rightarrow
\mathbb{R}$ which are of class $ C^{r}, r \geq 2 $. \\

In a next step the constitutive equations have to be specified. In
a theory of type $\Phi_{R} $ as treated in this paper there is
only one constitutive element, the energy-momentum tensor $ T $,
and the constitutive equation relates $ T $ to the structural
terms $ g, v, \eta, \vartheta $. This is the content of the next
axiom where the bundle of $ {n}\choose {m}$-tensors is denoted $ T_{m}^{n} $. \\

\textit{\bf EM2}: $ T $ is a function $T : M \rightarrow T_{0}^{2}
M$ which is of class $ C^{r}, r \geq 2 $ and which is defined by
the functional $ {\mathcal{T}} $ of $ g, v, \eta, \vartheta $ for
all $ q \in M $ by $T (q) = {\mathcal{T}} ( g, v, \eta, \vartheta)
( q) \in T^{2}_{0 q} M$.
If $ \eta (q) = 0$ for $ q \in N \subset M $ then ${\mathcal{T}}  ( g, v, \eta, \vartheta) ( q) = 0$.\\

This axiom is the point why $ \Phi_{R} $ is a frame theory, for $
{\mathcal{T}} $ is not explicitly specified. Each such
specification of $ {\mathcal{T}} $ defines a subclass of the
systems governed by the (frame) theory $ \Phi_{R} $. The usual
forms of the energy-momentum tensor for systems of dust or for
general Euler fluids are examples of classifying systems with the help of $ \mathcal{T}$. \\

The next two axioms are the equation of continuity and the balance
of energy and momentum. Though the latter is a consequence of
Einstein´s equation it is written down here. \\

\textit{\bf EM3}: Throughout $M: \mathrm{div} ( \eta v)=0$. \\

\textit{\bf EM4}: Throughout $M: \mathrm{div} (T)=0$.

\subsection{Einstein equation}

In our context there is no need to comment on Einstein's equation
or to motivate it. It suffices to write it down. So let, as usual,
$R$ be the Ricci tensor, $\bar{R}$ the Ricci scalar, $\Lambda_{0}$
the (unspecified) cosmological constant and finally $ \kappa_{0}$
Einstein´s gravitational constant. Moreover, $T^{\flat}$
denotes the covariant energy-momentum tensor. \\

\textit{\bf EE}: Throughout $M$: \quad $R - \frac{1}{2}g \bar{R} +
\Lambda_{0} g = \kappa_{0} T^{\flat}$.

\subsection{Additional conditions}

The general term additional conditions (AC) comprises all those
axioms which have to be imposed in order to get physically
relevant and uniquely determined classes of models. In this
context, by a model (or a ''solution'') the following is to be
understood:  \\

\textit{\bf Notation (4.1)}:  Let $ M', P', S'$  and $ \Psi', g',
v', \eta', \vartheta'$ be terms defined within mathematical
analysis or, more precisely, within the theory of sets such that
the axioms GK, EM, EE and AC for a specified functional $
\mathcal{T}$ and a specified cosmological constant $ \Lambda_{0} $
are satisfied. Then we say that these terms define an {\it
analytical} or {\it a set theoretical model}
of the frame theory $ \Phi_{R} $. \\

A model is a theory of the form described in (ii) of Section 1.2.
In this sense Robertson-Walker space-times, the Schwarzschild
space-time  etc. are models of $ \Phi_{R}$. But this is not
proved here. \\

There is a great variety of AC. Three examples may illustrate the
role AC play:
(i) Initial conditions, e.g.\ for solving a Cauchy problem. \\
(ii) Boundary conditions, e.g.\ for space-times which are to be
assymptotically flat in a certain sense. \\
(iii) Symmetry conditions by which e.g.\ a general ansatz can be
restricted to a more special form. \\

A more detailed discussion of this subject is outside the scope of
this paper. \\

\section{Some Consequences of the Axioms}

\subsection{Metric and velocity in B-coordinates}

 First of all let us fix some

 \textit{\bf Notation (5.1)}: The components of $v$ and $g$ with
 respect to B-coordinates (cf. Definition (3.2)) for each $B \in P$ are
 denoted

\begin{equation*}
v_{B}^{\alpha} (y) \quad \text{and} \quad g_{\alpha \beta}^{B}
{(y)} \quad \text{where} \quad y  \in \psi_{B} [V_{B}].
\end{equation*}

 For the general coordinates $ \chi$ with $ (W,  \chi) \in \mathcal{D}$ we write
\begin{equation*}
v_{\chi}^{\alpha} (x) \quad \text{and} \quad g_{\alpha
\beta}^{\chi} (x) \quad \text{where} \quad x \in \chi [W].
\end{equation*}

 Then the following simple lemma holds: \\

 \textit{\bf Lemma (5.2)}: For each $B \in P$ and for all $ y \in \psi_{B} [W_{B}] $ it holds that
\begin{equation*}
 v_{B}^{\alpha} (y) = \delta_{4}^{\alpha}.
\end{equation*}
{\bf Proof}: By Axiom GK5 and by making use of Definition (3.4)
 for all $b \in W_{B}$ one finds that $v (b)=\dot{\gamma_{B}} (t)$ where $t={\mathcal U}_{B} (b)$.
Let $\dot{\gamma_{B}} (t)=w_{B}^{\alpha} (t)
\partial_{\psi_B^\alpha} $. Then

\begin{equation*}
w_{B}^{\alpha} (t) = \frac{d}{dt} (\psi_{B}^{\alpha} \circ
\gamma_{B}) (t).
\end{equation*}

 From Remark (3.5) it follows that $w_{B}^{\alpha} (t) = \delta_{4}^{\alpha}$ for all $t= {\mathcal U}_{B}(b)$ and
 $b \in W_{B}$, i.e. for all $t \in J_{B}$. Let $y = \psi_{B}(b)$ and $t = {\mathcal U}_{B} (b)$. Then for each
 $b \in W_{B}$ we have $v_{B}^{\alpha}(y)=w_{B}^{\alpha}(t)$, so that the proposition is proved. \\

 A similar result holds for the metric $g$: \\

 \textit{\bf Lemma (5.3)}: For each $B \in P$ and for all $y \in \psi_{B} [W_{B}] $ it holds that
\begin{equation*}
g_{\alpha \beta}^{B} (y) = \eta_{\alpha \beta}
\end{equation*}

 where $((\eta_{\alpha \beta})): = \mathrm{diag} (1,1,1,-1)$ is the Minkowski matrix. \\

 {\bf Proof}: In what follows, for the sake of simplicity the
 argument $(y)$ is omitted. \\
 1. From the Axioms GK 3.4 and GK 5 it follows that $g_{\alpha \beta}^{B} v_{B}^{\alpha} v_B^\beta = -1$.

By Lemma (5.2) we have $v_{B}^{\alpha}= \delta_{4}^{\alpha}$ so
that $g_{\alpha \beta}^{B} v_{B}^{\alpha} v_{B}^{\beta} = g_{4
4}^{B}  = \eta_{4 4}$.

2. From the Axioms GK 4.1 and 4.2 one concludes that for all
$(e^{1}, e^{2},e^{3})$ the relation

\begin{equation*}
\sum_{j \textit{k}}^{3} g_{j \textit{k}}^{B} e^{j} e^{\textit{k}}
+ 2 \sum_{j}^{3} g_{j 4} e^{j} - 1=0
\end{equation*}

holds. Now let $ e^{1}= \pm 1, \; e^{2}=0, \;e^{3}=0$. Then
$g_{11}^{B} \pm 2g_{14}^{B} = 1.$
Hence $g_{11}^{B} = 1 = \eta_{11}, \quad g_{14}^{B}=g_{41}^{B}=0=\eta_{14}.$ \\
Similarly we find that $g_{22}^{B}=1= \eta_{22}, \quad
g_{24}^{B}=g_{42}^{B} = \eta_{24}, \quad g_{33}^{B}=1 =
\eta_{33},$ and $ g_{34}^{B} = g_{43}^{B} = \eta_{34}.$

Finally let $ e^{1}= \frac{1}{\sqrt{2}}, e^{2}=
\frac{1}{\sqrt{2}}, e^{3}=0  $. Then $g_{12}^{B} = g_{21}^{B} = 0
= \eta_{12} = \eta_{21}.$

By analogous arguments one finds that $g_{j k}^{B} = 0 = \eta_{j k}, \; k = 1, 2, 3,  \; j \not= k.$\\

\textit{\bf Remark (5.4)}: 1.  Lemma (5.3) is remarkable in so far
as it states that for each observer $B$  the metric in
$B$--coordinates, i.e.\ $g_{\alpha \beta}^{B}$ is Minkowskian not
only for one point but for each point of the whole worldline
$W_{B}$. Therefore, by Axiom GK 2.1 the metric $g$ as well as the
velocity $v$ is completely determined on $M$ once for all $B \in
P$ the worldlines $W_B$ are known. These in turn are
determined by the coordinate function $ \hat{\Psi}$ resp.\ $ \Psi$. \\
2. As already mentioned in Section 1.3, the proper radar
coordinates are also pre-radar coordinates, i.e.\ B-coordinates for
some observer B. Also the Fermi coordinates introduced by Synge in
\cite{Synge71} p.\ 84 can be used to define pre-radar coordinates.
But there
are pre-radar charts which are neither radar charts nor Fermi charts.\\
3. It is stated without proof that a pre-radar observer $B$ is freely falling exactly if

\begin{displaymath}
\frac{\partial}{\partial y^j} g^{B}_{44}  =  0
\end{displaymath}

for all $y \in \psi_{B}[W_{B}] $ and j = 1, 2, 3.

\subsection{Coordinate Representations for metric and velocity}

In this section we are looking for explicit expressions for
$g^{\chi}_{\alpha \beta}$ and $v_{\chi}^{\alpha}$ where $ \chi$
are arbitrary
coordinates. First of all we have to fix some notation. \\

{\bf Definition (5.5)}: Let $(W, \chi) \in \mathcal{D}$ and
$(V_{B},\psi_{B}) \in \mathcal{A}$. Then:\\[0,1cm]
1. $ \phi_{\chi B}: = \psi_{B} \circ \chi^{-1}, \; \phi_{B \chi} := \phi_{\chi B}^{-1},  \;  \phi_{A B }: = \psi_{B} \circ \psi_{A}^{-1}$.\\[0,1cm]
2. $ G_\chi: = F (\chi^{-1})$, where $F$ is defined in Remark (3.6).\\[0,1cm]
3. $ \phi_{\chi} : = \Psi (\chi^{-1}, \chi^{-1})$ (cf. Definition (3.7)).\\[0,1cm]
4. $\Lambda_{ \chi} = ((\Lambda_{\chi}\left._
\beta^\alpha\right.))$ with
$\Lambda_\chi\left._\beta^\alpha\right. (x):= \left.
\frac{\partial \phi^\alpha_\chi (x, z )} {\partial
x^{\beta}}\right|_{z=x}$
where $\alpha$ denotes the rows and $\beta$ the columns. \\

\textit{\bf Remark (5.6)}: Since $ \Psi(p,q)=\psi_{F(q)} (p) $ it
follows that

\begin{equation*}
\phi_{ \chi}(x, z) = \psi_{G_{ \chi}(z)} ( \chi^{-1} (x)) =
\phi_{\chi G_{\chi} (z)} (x).
\end{equation*}

Next the representation for the metric is deduced. \\

{\textit{\bf Proposition (5.7)}}: If $(W,\chi) \in \mathcal{D} $
then for all $ x \in \mathrm{ran} \chi$ we have

\begin{equation} \label{prop5.7}
g^{\chi}_{\alpha \beta}  (x) =
\Lambda_\chi\!\left.^{\kappa}_\alpha\right.(x)
 \Lambda_\chi\!\left.^{\lambda}_{\beta}\right.(x) \eta_{\kappa \lambda}.
\end{equation}

{\bf Proof:}  Let $ x = \chi (p)$ and $ y= \psi_{B}(p)$. Then $y =
\phi_{\chi B} (x)$. The matrix elements $g^\chi_{\alpha \beta}(x)$
and $ g^B_{\kappa \lambda}(y)$ are related by

\begin{equation*}
g^{\chi}_{\alpha \beta} (x) = \frac{\partial{\phi}^{\kappa}_{\chi
B} (x)} {\partial x^{\alpha}} \quad
\frac{\partial{\phi}^\lambda_{\chi B} (x)} {\partial x^{\beta}}
\quad g^B_{\kappa \lambda}(y).
\end{equation*}

Therefore, if $B= G_{\chi}(z)$ it follows from Remark (5.6) that

\begin{equation*}
g^\chi_{\alpha \beta} (x) = \frac{\partial{\phi^\kappa_{\chi} (x,
z)}} {\partial x^{\alpha}} \frac{\partial{\phi}^{\lambda}_{\chi}
(x,z)} {\partial x^{\beta}} g^B_{\kappa \lambda}(y).
\end{equation*}

Now let $z=x$. Then $B= G_{\chi}(x) = F(\chi^{-1}(x)) = F(p)$, so
that $p \in W_B$ and $y \in \psi_B[W_B]$. Hence by Lemma (5.3) we
have $g^B_{\kappa \lambda} (y) = \eta_{\kappa \lambda}$ for all $y
\in \psi_B[W_B]$,
so that by use of Definition (5.5) the proposition holds.\\
Also the components of velocity $v$ can be expressed by $\Lambda_\chi$ in the following way.\\

{\textit{\bf Proposition (5.8):}} If $(W,\chi) \in \mathcal{D}$
then for all $ x \in \mathrm{ran} \chi$ we get:

\begin{equation} \label{prop5.8}
v^\alpha_\chi (x) = \Lambda^{-1}_\chi\! \left.^\alpha_4\right. (x)
.
\end{equation}

{\bf{Proof:}} Let $x = \chi(p), \quad y = \psi_B(p) \quad
\text{and} \quad x = \phi_{B\chi}(y)$. Then

\begin{equation*}
v^\alpha_\chi (x) = \frac{\partial \phi^\alpha_{B \chi}(y)}
{\partial y^\beta} \quad v^\beta_B (y).
\end{equation*}

By Definition (5.5) we have $\Phi_{B \chi} = \Phi^{-1}_{\chi B}.$
Therefore

\begin{equation*}
v^\alpha_\chi (x) = \left[ \left( \frac{\partial \phi_{\chi B}
(x)} {\partial x}  \right)^{-1}\right]^\alpha_\beta v_B^\beta (y).
\end{equation*}

If $B=G_\chi(z)$ then from Remark (5.6) it follows that

\begin{equation*}
v^\alpha_\chi (x) = \left[\left( \frac{\partial \phi_{\chi} (x,z)}
{\partial x}  \right)^{-1}\right]^\alpha_\beta v_B^\beta (y).
\end{equation*}

Finally let $x=z$. Then $B = G_\chi(x) = F(p)$, hence $p \in W_B$
and $y \in \psi_B[W_B]$. Therefore, from Lemma (5.2) we have
$v^{\alpha}_{B}(y) = \delta^{\alpha}_{4}$.
Using Definition (5.5) the proposition is proved. \\

{\bf{Corollary (5.9):}} For the covariant components of the
velocity the relation

\begin{equation} \label{corol5.9}
v_\alpha^\chi (x) = - \Lambda_\chi \!{ \left.^4_\alpha\right.}
(x).
\end{equation}

holds. For using the Propositions (5.7) and (5.8), we have

\begin{equation*}
v_\alpha^\chi =  \Lambda^{-1}_\chi \left.^\beta_4 \right.
\Lambda_\chi\!{\left.^{\kappa}_{\alpha}\right.} \,
\Lambda_\chi\!\left.^\lambda_\beta \right. \, \eta_{\kappa
\lambda} = -  \Lambda_{\chi}\!\left.^4_\alpha\right. .
\end{equation*}

Similarly, the contravariant components of the metric are

\begin{equation} \label{corol5.9-1}
g_{ \chi}^{ \alpha \beta} =
\Lambda^{-1}_\chi\!\left.^\alpha_\kappa\right.
 \Lambda^{-1}_\chi\!\left.^\beta_\lambda\right. \eta^{\kappa \lambda}.
\end{equation}

This follows directly from $((g^{\alpha \beta}_{\chi})) =
((g_{\alpha \beta}^{\chi}))^{-1} \; \text{and} \;  \eta^{\kappa
\lambda} = \eta_{\kappa \lambda}.$

\subsection{Global representations}

In this section the results of Section 5.2 will be shaped in a
form which is independent of coordinates.
These considerations show again the role the function ${\Psi}$ plays. We start with some notation.\\
{\bf{Definition (5.10):}} Let $\Psi$ be the function introduced in
Definition (3.7). Then

\begin{eqnarray}
e_{q \alpha} (p) : = \left. \partial_{\Psi^\alpha ( \cdot ,q)}
\right|_p , \\ \label{def5.10} \Theta^\beta_q (p) : = \left. d
\Psi^\beta (\cdot ,q)  \right|_p.   \label{def5.10-1}
\end{eqnarray}

{\textit{\bf Remark (5.11):}} Since $\Psi (\cdot ,q)$ is a
coordinate function, $(e_{q1}(p), \ldots , e_{q4}(p))$ is a tetrad
in $T_p M$ and $(\Theta_q^1(p), \ldots , \Theta_q^4(p))$ is the
dual tetrad in $T_p^\ast M$. Hence for all $p \in V_{F(q)}$ we
have:

\begin{equation}   \label{remark5.11}
\Theta_{q}^{\beta} (p) (e_{q \alpha}(p)) = \delta_{\alpha}^{\beta}
= e_{q \alpha}(p)(\Theta_{q}^{\beta} (p)).
\end{equation}

In what follows we need only a special form of these bases.\\

{\textit{\bf Notation (5.12):}} 1. For all $p \in M$ we write:

\begin{equation} \label{not5.12}
e_{\alpha} (p) := e_{p \alpha} (p) \, , \, \Theta^\beta (p) :=
\Theta_{p}^{\beta} (p).
\end{equation}

2. To simplify notation, the arguments $(p), (x)$ etc. and the
index $\chi$ indicating a coordinate system are mostly omitted in
the
sequel.\\

{\textit{\bf Proposition (5.13):} Let $(W, \chi) \in \mathcal{D}$.
Then for all $p \in W, x= \chi(p) \; \text{and} \; \Lambda =
\Lambda_{\chi}$:

\begin{eqnarray} \label{prop5.13}
\Theta^{\beta}(p) &=& \Lambda^{\beta}_{\kappa}(x) dx^{\kappa}, \\
e_{\alpha}(p) &=& \Lambda^{-1}\! \left.^\lambda_\alpha \right. (x)
\partial_{x^\lambda}.  \label{prop5.13-1}
\end{eqnarray}

{\bf Proof:}} Let $q$ be fixed and $z = \chi(q)$.  Then $y =  \Psi
( \chi^{-1}(x), q ) = \phi (x,z)$ is the transformation between
$\chi$-coordinates $x$ and pre-radar coordinates $y$. Therefore

\begin{equation*}
dy^{\beta} = \frac{ \partial \phi^{\beta}}{ \partial x^{\kappa}}
dx^{\kappa} , \quad \partial_{x^\alpha} =  \frac{\partial
\phi^{\lambda}}{ \partial x^{\alpha}}
\partial_{y^{\lambda}}.
\end{equation*}

Hence

\begin{equation*}
\partial_{y^{\lambda}} = \left[\left( \frac{\partial \phi}{\partial x} \right)^{-1}\right]^ \alpha_{\lambda} \partial_{x^\alpha}.
\end{equation*}

With $z=x$ and Definition (5.5) the proposition is seen to hold. \\

{\bf Remark (5.14):} The matrix elements $\Lambda^\beta_\kappa,
\kappa = 1, \ldots, 4$ are the $ \chi$-components of
$\Theta^\beta$ and the
$\Lambda^{-1}\left.^\lambda_{\alpha}\right., \lambda = 1, \cdots
4$ are the $\chi$-components of $e_\alpha$, i.e.
$\Lambda^\beta_\kappa = \Theta^\beta (\partial_{x^\kappa})$ and
$\Lambda^{-1}\!\left.^\lambda_\alpha\right. = e_\alpha (d
x^\lambda)$. Hence under transformation of coordinates they
transform like components of covectors and vectors.

{Proposition (5.13)} has an immediate consequence for $g$ and $v$.
Inserting the formulae (\ref{prop5.13}) and (\ref{prop5.13-1})
into formulae (\ref{prop5.7}) to (\ref{corol5.9-1}) we arrive at the following result.\\

{\bf Proposition (5.15):} Throughout $M$ we have:

\begin{equation} \label{prop5.15}
g = \eta_{\alpha \beta} \Theta^\alpha \otimes \Theta^\beta, \quad
g^\sharp = \eta^{\kappa \lambda} e_\kappa \otimes e_\lambda, \quad
v^{\flat} = - \Theta^4, \quad  v = e_4.
\end{equation}

Therefore, it is justified to say that $\Psi$ generates $g$ and  $v$ .\\
The result can be stated thus: with respect to
$(\Theta^1,.....,\Theta^4)$ the tetrad components of $g$ are
$\eta_{\alpha \beta}$, and with
respect to $(e_1,.....,e_4)$ the tetrad components of $v$ are $\delta_{4}^{\alpha}$.\\

{\bf Remark (5.16):} Using Proposition (5.15) and Equation
(\ref{remark5.11}) we find the orthogonality relations

\begin{equation} \label{rem5.16}
g (e_\kappa, e_\lambda) = \eta_{\kappa \lambda} , \quad g^\sharp
(\Theta^\alpha, \Theta^\beta) = \eta^{\alpha \beta},
\end{equation}
and
\begin{equation} \label{rem5.16-1}
\eta^{\kappa \lambda} g(e_{\lambda},  \cdot) = \Theta^\kappa,
\quad  \eta_{\alpha \beta} g^\sharp (\Theta^\beta, \cdot) =
e_\alpha.
\end{equation}

Roughly speaking, the result of chapter 5 is this: the function
$\Psi$ is a "potential" such that the metric $g$ and the velocity
$v$ are determined by the derivates of $\Psi$. Moreover, $\Psi$
itself has a physical meaning, namely $\Psi (\cdot, q)$ is a
coordinate function for each $q \in M$.

The existence of the fields $\Theta^\alpha, e_\beta, \alpha, \beta
= 1, \cdots, 4,$ hence the existence
of the function $\Psi$ which determines $\Theta^\alpha$ and $e_\beta$, has a consequence concerning time:\\

{\bf Remark (5.17):} 1. From the Axioms GK 3.4 and GK 5 and from
Proposition (5.15) one concludes that $e_4 = v$ is a timelike
$C^r$-vector field, $r \ge 2$ on $M$ so that it nowhere vanishes.
Therefore, the space-time manifold $(M, {\mathcal{A}}, g)$ is time
orientable (cf.\ e.g.\ \cite{sachs77} p.\ 26). Already from
Definition (3.4) where the clock  ${\mathcal{U}}_A$ of a particle
$A \in P$ is introduced we conclude that $(W_A, {\mathcal{U}}_A)$
is a one-dimensional manifold with a global chart for each $A \in P$. Therefore, $W_A$ cannot be a closed curve.\\
2. Following Geroch \cite{Geroch68}, in a noncompact space time
the existence of a smooth global field of tetrads is a necessary
condition of a spinor structure.

{\bf Remark (5.18):} The function $\Psi$ determines not only $g$
and $v$ but also a worldfunction $\Omega$ (cf. \cite{Synge71}) by

\begin{displaymath}\Omega (p, q) = \eta_{\kappa \lambda} ( \Psi^{ \kappa}(p, q) -
\Psi^{ \kappa}(q, q)) (\Psi^{\lambda}(p, q) -
\Psi^{\lambda}(q,q)).
\end{displaymath}

Then it is easily seen that for $\bar{\Omega}(x, z) :=
\Omega(\chi^{-1}(x), \chi^{-1}(z))$ the relation

\begin{displaymath}
\frac{\partial}{\partial x^{\alpha}} \frac{ \partial}{\partial x^{
\beta}} \bar{ \Omega}(x, z) \big|_{z=x}  =  g^{ \chi}_{ \alpha
\beta}
\end{displaymath}

holds. The same result is obtained if one takes the covariant
derivatives.

\section{Alternative Axiomatics}

In the investigation so far the set $S$ of signals was introduced
solely to ensure that Lemma (5.3) is provable. Thus the
question arises if it is possible to avoid $S$ by changing some axioms. This is indeed the case. To see this, let us introduce some\\

 {\bf Notation (6.1):} Let $\Phi^{\ast}_R$ be the frame theory which has the base sets $M, P, \mathbb{R}$ and the structural terms
$\hat{\Psi}, g, v, \eta, \vartheta$ which are subject to the
following axioms: GK 1.1 to 1.4, GK 2.1 to 2.3, EM, EE and AC
together with GK*3 and 4 which read:\\

{\bf GK*3:} For all $p  \in M$:

\begin{equation*}
g (p) = \left. \eta_{\alpha \beta} \; d \Psi^\alpha (p, q)
\right|_{q = p}  \otimes d \Psi^\beta (p, q) |_{q = p}.
\end{equation*}
{\bf GK*4}: For all $p \in M:  v(p) = \partial_{\Psi^{4}(p, q)} \big|_{q=p}$,\\

Then the following theorem holds.\\

{\bf Proposition (6.2):} The theories $\Phi_R$ and $\Phi^\ast_R$
are equivalent in the following sense. The theory $\Phi_R$ is
stronger than $\Phi^\ast_R$, i.e.\ all axioms of $\Phi^\ast_R$ are
theorems in $\Phi_R$. Conversely, there is a term $S^\ast$ defined
in
$\Phi^\ast_R$ such that all axioms of $\Phi_R$ are theorems in $\Phi^\ast_R$ if the letter $S$ is replaced by the term $S^\ast$.\\

{\bf Proof:} From Proposition (5.15) one concludes that $\Phi_R$
is stronger than $\Phi^\ast_R$. To see the converse, one has to
show that the Axioms GK 3 to GK 5 of $\Phi_R$ are theorems in
$\Phi^\ast_R$. The term $g$ in Axiom GK$^\ast$ 3 is a $0\choose
2$-tensor field throughout $M$ and $g (p)$ acts on $T_p M \times
T_p M$ for all $p \in M$. Moreover, $g$ is of class $C^r, r \ge
2$, because $\Psi$ is of class $C^k, k \ge 3$. Finally $g$ is
symmetric, non-degenerate and of signature 2 because $\hat{\eta}
=$ diag (1,1,1, -1) has these properties. Therefore, the axioms GK
3.1 to 3.3 are theorems in $\Phi^\ast_R$.

Within $\Phi^\ast_R$ the function $\gamma_A$ is defined like in
$\Phi_R$. Consequently, by Axiom GK*4 together with Definition
(5.10) and Notation (5.12) one has $v(p) = e_{4}(p)$. Moreover,
from Definition (3.3) it follows that $\dot\gamma_A(t) =
e_4(\gamma_A(t))$. Therefore,

\begin{equation} \label{prop6.2}
v(p)  =  \dot\gamma_{F(p)} (t),\quad   t  =  \gamma^{-1}_{F(p)}(p)
\end{equation}
so that Axiom GK 5 holds in $\Phi^\ast_R$. Because of
$\Theta^\alpha (e_4) = \delta^\alpha_4$ also
GK 3.4 is a theorem in $\Phi^\ast_R$.\\

In order to prove the Axioms GK 4.1 and 4.2 in $\Phi^\ast_R$ one
has to define a term $S^\ast$ so that these axioms can be verified
in $\Phi^\ast_R$ if $S$ is replaced by $S^\ast$. The definition of
$S^\ast$ runs as follows. For each $A \in P$ let us consider the
set of all null geodesics $\sigma$ which start at $W_A$, and let
$\zeta = \psi_A \circ \sigma$. Then $\sigma$ obeys the equations

\begin{equation} \label{proof6.2}
\nabla_{\dot \sigma}  \dot\sigma = r \dot \sigma
\end{equation}

and 

\begin{equation}  \label{proof6.2-1}
g (\dot\sigma, \dot\sigma) = 0.
\end{equation}

Let the parameter of $ \sigma$ be denoted $\lambda$. Then in
$\psi_A$-coordinates let $\zeta (\lambda_0) = x$, where $x =
\psi_A (p)$, $p \in W_A$. (Generally $\lambda_0$ depends on $p!$)
Now for each $p \in W_A$, (6.3) reads

\begin{equation} \label{proof6.2-2}
\eta_{\alpha \beta} \dot{\zeta}^\alpha (\lambda_0)
\dot{\zeta}^\beta (\lambda_0) = 0.
\end{equation}

Hence $\dot{\zeta}^4 (\lambda_0) > 0$ because only starting null
geodesics are considered. This means we can choose $t = \zeta^4
(\lambda)$ as a new parameter. Thus, using the same symbol
$\zeta$, we have $\dot\zeta^4 (t_0)= 1$. From (6.4) it follows
that $\dot\zeta (t_0) = (e^1, e^2, e^3, 1)$  where $(e^1, e^2,
e^3) \in S^2 (S^2$ being the 2-sphere). Now we define $\hat{S}_A$
to be the set of all null geodesics $\sigma$, i.e. solutions of
(\ref{proof6.2}) and (\ref{proof6.2-1}) which in
$\psi_A$-coordinates obey the initial conditions $\zeta (t_0)  =
x_0, \quad t_0 = x^4_0 \quad \text{for any} \quad  x_0 = \psi_A
(p), \quad p \in W_A$ and $\dot\zeta (t_0) = (e^1, e^2, e^3, 1)$
for any $(e^1, e^2, e^3) \in S^2$.

Finally, let $\hat{S} = \cup_{A \in P}  \hat{S}_A$ and let
$S^\ast$ be any set of the same cardinality as $ \hat{S}$ which
may serve as a set of indices for $\hat{S}$.

Then it is easily seen that the Axioms GK 4 are satisfied if $S$ is replaced by $S^\ast$.\\

This axiomatics is the simplest one for the considered systems.

\vfill
{\bf Acknowledgement}

I want to thank Mr. Gerhard Lessner for valuable discussions and
critical remarks and Mr. Wolfgang Rothfritz for correcting my
English.% \clearpage

\bibliographystyle{my-h-elsevier}

\end{document}